# Evolving Military Broadband Wireless Communication Systems: WiMAX, LTE and WLAN

P. Fraga-Lamas
and L. Castedo-Ribas
Department of Electronics and Systems
University of A Coruña, Spain
Email:paula.fraga@udc.es, luis.castedo@udc.es

A. Morales-Méndez
and J. M. Camas-Albar
Indra Sistemas, S.A.
Aranjuez, Spain
Email: ammendez@indra.es, jmcamas@indra.es

*Abstract*—Emerging technologies for mobile broadband wireless are being considered as a Commercial Off-The-Shelf solution to cover the operational requirements of the future warfare. The capabilities of these technologies are being enhanced to meet the growing market demands on performance. In this context, several standards such as WiMAX, LTE or WLAN are introducing themselves as strong candidates to fulfill these requirements. This paper presents an innovative scenario-based approach to develop a Military Broadband Wireless Communication System (MBWCS). Its main objective is to analyze how similar a military MBWCS can be to the identified civil standards, taking operational and high level technical requirements into account. This specification will be used for analyzing the applicability and the modifications of each of the standards layers individually. Proving the feasibility and aptitude of each standard provides strong foundations to address a MBWCS in the most efficient way.

## I. Introduction

The motivation to develop military disruptive technologies is driven by new operational needs and the challenges arising from modern military deployments. The fast evolution of Commercial Off-The-Shelf (COTS) technologies is one of the primary reasons to analyze the usage of these up-to-date technologies to fulfill current tactical deployment necessities.

This study focuses on the strategic advantages of broadband technologies massively deployed in civil scenarios, such as 4G Worldwide Interoperability for Microwave Access (WiMAX), Long-Term Evolution (LTE) and Wireless Local Area Network (WLAN).

Under these assumptions, the military Data Distribution Subsystems (DSS) have the greatest similarity with commercial wireless technologies in terms of communication range, requested services and network capabilities support. Nevertheless, it is not possible to use these technologies due to the specific characteristics of tactical environments.

This scenario-based approach together with a deep analysis of modern civilian waveforms is a guarantee to minimize the impact and cost of a new Military Broadband Wireless Communication System (MBWCS) development. Current market COTS 4G-based tactical products and on-going international waveform development initiatives, such as COALWNW, ESSOR, NATO Narrowband WF and other MANET/Ad Hoc were also examined to confirm that a 4G-based MBWCS can coexist, and that it makes sense to devote effort to its definition and development.

The aim of our scenario-based approach is to determine the technologies required in the middle and long term to comply with the operational requirements, and the state-of-the-art COTS military equipment that covers such needs. After the definition of the NATO scenarios, an analysis of the operational requirements is performed. In a second step, the technical requirements are derived and used as input for the applicability analysis. For this work, it was necessary to characterize the technical implementation requirements of 4G standards, and analyze capabilities, aptitudes and challenges of deploying a tactical network. Also, modifications and its related techniques related to the three standards are identified and evaluated.

The context of this study is framed within NATO Exploratory Team IST-ET-068: 'LTE vs. WiMAX for Military Applications'. This ET was launched to analyze the applicability of the deployed 4G wireless standards in a tactical environment. The main objective of the ET has been to assess whether it is worth adapting high-performance 4G standards or it is better to develop a new system from scratch in order to cover imminent demands in the tactical domain. We have relied on previous and on-going NATO research: Cognitive Radio (I and II: IST-104-RTG-035 & RTG-055, SDR (IST-080), Military Communications and Networks (IST-092), Tactical Communications in Urban Operations (IST-067), Emerging Wireless Technologies (IST-070) and Next Generation Communications (IST-105), trends on this field and the background of the different partners in national and international initiatives.

The remainder of this article is structured as follows. Section 2 provides a brief overview of the state-of-the-art of WiMAX, LTE and WLAN broadband wireless standards. Section 3 characterizes the relevant scenarios in which these technologies may be applicable within NATO countries tactical deployments. Next, the operational requirements and end-users needs that will drive the technical analysis are explained. The methodology and the applicability analysis results are reflected in Section 5 and 6. Finally, Section 7 is devoted to the conclusions, future research lines and to foster a roadmap for implementation.

## II. Standards overview

Commercial broadband cellular technologies offer high value for situation awareness, monitoring and intervention,

TABLE I: Comparison between WiMAX, LTE and Wi-Fi.

| Metric | WiMAX 2 (IEEE 802.16m) | LTE-A (3GPP Rel 10) | Wi-Fi (IEEE 802.11n) |
|---|---|---|---|
| Technology orientation | Flat All-IP architecture initially born as Fixed WiMAX. Data-oriented evolved to support voice. | Focused in voice, progress gradually for data services (GSM/GPRS/EGPRS/UMTS/HSPA). | |
| Frequency bands | LOS: 10-66 GHz NLOS: 2-11GHz licensed and unlicensed bands | 700, 1700, 1900, 2100, 2500 and 2600 | 2.4 and 5 GHz |
| FFT Size | 1.25 MHz to 28 MHz / 128 - 2048 | 128 - 2048 | 20 MHz or 40 /64 or 128 |
| Physical layer | DL/UL: OFDMA | DL: OFDMA, UL: SCFDMA | OFDM (200 channels) |
| Duplex mode | TDD, FDD and H-FDD | TDD, FDD (originally more interest in FDD) | TDD |
| Modulations | QPSK, 16-QAM and 64-QAM | QPSK, 16-QAM and 64-QAM | BPSK, QPSK, 16-QAM and 64-QAM |
| Mobility | Max 350 km/h | Max 350 km/h | 200 km/h (IEEE 802.11p) |
| Coverage | Up to 50 km | Up to 100 km | > 200 m |
| Operating bandwidth | 5, 7, 8.75, 10, 20, and 40 MHz (up to 100 MHz with carrier aggregation). | Up to 100 MHz | 5, 10, 20 and 40 MHz |
| Peak data rate | DL: >350 Mbps (MIMO 4×4), UL: >200 Mbps (MIMO 2×4) with 20 MHz and FDD | DL: 1 Gbps, UL: 500 Mbps | 6-600 Mbps (MIMO 4×4) |
| Average cell spectral efficiency | DL: >2.6 bps /Hz (MIMO 2x2), UL: >1.3 bps/Hz (MIMO 1x2) | DL: >1.6-2.1 bps /Hz, UL:> 0.66-1 bps/Hz | >3bps/Hz |
| Latency | Link layer < 10 ms, Handover < 30 ms | Link layer < 5 ms, Handover < 50 ms | Handover < 50 ms (IEEE 802.11f and 802.11r) |
| Security | WPA2 | WPA2 | WPA2 (802.11i) |
| VoIP capacity | >30 users per sector /MHz (TDD) | >80 users per sector /MHz (FDD) | 12 active calls IEEE 802.11 a/b/g/n |
| Additional features | QoS | QoS | QoS (IEEE 802.11e), Dynamic Frequency Selection and Transmit Power Control (IEEE 802.11h) |
| Roadmap | IEEE 802.16-2012 (Revision of IEEE 802.16 including Std 802.16h, IEEE Std 802.16j y IEEE Std 802.16m (WirelessMAN-Advanced is part of IEEE Std 802.16.1). IEEE 802.16p (First Amendment to IEEE 802.16-2012), M2M applications. IEEE 802.16n (Second Amendment to IEEE Std 802.16-2012), Higher Reliability Networks. IEEE 802.16q (Third Amendment to IEEE Std 802.16-2012), Multi-tier Networks. | Rel-11, 2013 (CoMP, eDL MIMO, eCA, MIMO OTA · · · ). Rel-12, 2015 (new type of subcarrier, active antenna systems, ProSe, PTT, eMBMS). Rel-13, 2016 (LTE in unlicensed spectrum with Licensed-Assisted Access (LAA), Carrier Aggregation up to 32 component carriers and hence provide a major leap in the achievable data rates as well as flexibility to aggregate large numbers of carriers in different bands, enhancements for MTC, full-dimension MIMO, indoor positioning · · · | IEEE 802.11aa-2012 (MAC Enhancements for Robust Audio Video Streaming). IEEE 802.11ad-2012 (Enhancements for Very High Throughput in the 60 GHz Band). IEEE 802.11ae-2012 (Prioritization of Management Frames). IEEE 802.11ac-2013 (Enhancements for Very High Throughput for Operation in Bands below 6 GHz). IEEE 802.11af-2013 (Television White Spaces (TVWS) Operation). IEEE 802.11ad-2014 (transfer rate up to 7 Gbps). |

distributed command and control, and public participation in crisis management. WiMAX [1], LTE [2] and WLAN [3] are representative although competing technologies. Hence, there is a WiMAX-versus-LTE-versus-WLAN controversy to declare which one is the best. From a military point of view, there is a need to address which one, or which parts of them, best fits the operational requirements and target tactical deployments, but ignoring business related issues. These mainstream technologies resemble each other in some key aspects including scalable bandwidth, seamless mobility, operating in licensed spectrum bands, strong Quality of Service (QoS) mechanisms, and pure IP architecture. However, these technologies have evolved from different origins and differ from each other in certain aspects such as design choices, architecture, protocol stacks, air interface and security, as it can be seen in Table I.

## III. DEFINITION OF TARGET SCENARIOS

The final objective of our approach to NCW/NEC is to increase interoperability among networks compliant with the NATO NEC Feasibility Study recommendations, national operational needs and the proposed 'scenario based' methodology. Five main target land scenarios were identified:

**Type A: Battalion & Brigade level communication.**
This scenario can be defined as wireless communications between several Command and Control (C2) centers at battalion level and a C2 at Brigade level (also between two Brigade C2s or even division). The radius of action of Battalions is around 60 km, while the radius of action of Brigades will be approximately 150 km. Brigades can be composed of 4-20 battalions. Maximum distance in just one hop between CCs is approximately 50 km. It is a Line Of Sight (LOS) environment with no mobility and no need for MANET functionality on one side and a 100-150 Km single-hop range with mobility and a mesh scheme on the other one.

**Type B: Company & Battalion level communication.**
This scenario considers the provision of wireless communications between several C2 centers at Company and Battalion level. The environment fits in a typical rural environment with no significant obstacles and almost LOS between the different elements of the communication network. The maximum range of a Company is about 20 km, while at Battalion level is 60 Km. Battalions may be composed of 3-15 companies and the maximum distance in a single hop between C2 will be around 20 km. Mobility will be considered at both hierarchy levels. A mesh communication scheme would be adequate, i.e. a Command Center (CC) at Company level may contact with battalion level through other Company CCs within the range limit of communication.

**Type C: Wireless communication infrastructure at Battalion or Command HQ.**
This scenario covers a wireless communication infrastructure inside a Command Post to substitute traditional optical fiber deployments. It is typically a rapid deployment at Battalion HQ or Command Post (CP), equivalent to NATO Battalion CC. Hence, MBWCS technology can be deployed with fixed infrastructure allowing coverage within a radius of 2 km. The level of deployment risk and subsequent enhancements to existing COTS technologies will be negligible.

**Type D: Company level communications with limited mobility.**
This scenario can be defined as wireless communications to support Company CP communications (equivalent to a forward operating base). Fixed infrastructure with no or limited mobility is supported either via vehicles serving as a central access point to the network with antenna masts that can be elevated to maximize coverage, or through a deployable aerostat with a COTS access point. Typical coverage will be around 5 km. It is expected that the deployment risk will be increased to accommodate enhanced security and robustness.

**Type E: Full mobility Company level communications.** This scenario considers wireless communications with platoon deployment or Company/coalition dividing forces. In this scheme, a group can leave a fixed infrastructure network and form an ad-hoc MANET. In addition, robustness to interference and security issues will be key requirements. It is expected that this type of network will require a significant deployment risk while allowing the most flexible configuration of existing COTS products.

## IV. Operational requirements

A given set of operational requirements grouped by capabilities are presented in order to cover the previous scenarios.

### A. Deployment features

The MBWCS shall be a part of a military data network which enables integration of Command, Control, Communications, Computers, Intelligence, Surveillance and Reconnaissance (C4ISR) systems. The deployment will depend largely on the hierarchy of the unit. Large units shall have a semi-static or static character with non-restrictive time deployment (in an order of magnitude of hours). Small units will contemplate full mobility with rapid deployment (less than 10 minutes). Regarding the intrinsic features, MBWCS will be within determined ranges in terms of dimension, weight, heat dissipation and power consumption. In scenarios C, D and E, MBWCS target platforms will be portable, easily installable and dismountable. Except for C, hostile environments will be expected.

### B. System management and planning

The MBWCS will provide a simple GUI to enable easy network planning. It will include various user profiles to offer a selection of deployment features adapted to the user requirements. System management will be configured at Brigade level with the option of limited configuration at lower levels. MBWCS will support the ability to decentralize system management functions, plug and play capabilities with auto-configuration, and local and remote network management (in scenarios similar to Type E, where MANET functionality is required). System management will allow an ad-hoc network to form and separate from the existing network and rejoin an existing fixed infrastructure network, i.e. Type D scenario.

### C. Supported services and applications

The most critical and priority service is voice communication. In this way, Companies and Brigade and Battalion CCs will provide at least a verbal communication with low bandwidth and services like Push-To-Talk (PTT). Voice will take always priority over any other type of traffic; instant messaging, critical data and C2 messages, i.e. Blue Force Tracking (BFT). On the other hand, some important tactical data services, such as operation orders, fire support plans, logistics reports, cryptographic keys, configuration files as well as e-mails are also transferred between Brigade and Battalion CCs as well as between Battalion CCs and Companies. Nodes will be able to use IP based military applications such as C2, Combat Management System applications, ILS, surveillance and intelligence applications (map based applications, database lookup, etc). Some rules and parameters will be defined by the state-of-the-art QoS policies as well as by service prioritization mechanisms.

### D. Network capabilities

NATO Network Enabled Capability (NNEC) is enabled with Network Information Infrastructure (NII) to exchange timely and secure information between users from different NATO nations. The MBWCS shall support soft handover network mechanisms to support reliable communication in Type B, C and E scenarios where mobility is assumed. For scenarios A and D, no handover is needed.

The MBWCS will forward information through the network even when the range between communication nodes exceeds the coverage range. The network will adapt the transmission delay for optimization of QoS support.

### E. Supported network topologies

Military networks meet Command, Control, Communications, Computers, and Intelligence (C4I) system requirements facing the moves of users from one network to another or from one access interface to another. This implies the adaptation of the routing and maybe of the addressing. Back-up networks and reconfiguration functions will keep a maximum level of connectivity with adequate QoS. An IP-based, high-speed, extensible and reliable wireless tactical network will be established among land platforms. Connectivity requirements of nodes can be categorized as vertical communications up and down the command chain, horizontal between each level, horizontal at each level between adjacent formations, and horizontal and vertical outside the chain of command.

Network architecture primarily addresses Point-To-Multipoint (PMP) or Point-to-Point (PtP) links. These topologies are required in some of the scenarios identified above. Nevertheless, most platforms are mobile, and there is no chance of providing a communication infrastructure among them during the military operations. Therefore, MBWCS should be capable of establishing high-throughput ad-hoc networking for specific scenarios, i.e. MANET is required for small units (Type C, D and E). The mobile ad-hoc network is specially useful in rapid deployments. In PMP deployments used for small and big units, it is usual to require equipment that can aggregate four links. Fully mesh capabilities with network auto discovery, efficient automatic routing are critical at the small units. Relaying capacities can be used for range extension at the same hierarchical level, and between hierarchical levels operating at different frequency bands i.e. between Brigade and Companies. Network topology sizing will depend on the scenario and the level of hierarchy of the unit deployed. A reasonable assumption is 23 users per base station for the specific scenarios A and B, while a low number, from 5-15 users, will be necessary in scenarios C, D and E. When operating under Emissions Control (EMCON) restrictions, cooperative communications will not be possible.

*F. Mobility capabilities*

Brigade typically lacks mobility and presents a fixed infrastructure. On the other hand, Battalion and Company are mobile communication nodes. Land vehicles speed can change from 65 to 150 km/h. For Battalion CCs, the maximum speed can be considered around 100 km/h and the Armored Combat Vehicles (ACV) used for Company CCs around 150 km/h. A manpack radio or hand-held system can be used by a Company soldier or below in the field to join network with speeds up to 5 km/h. Close helicopter support shall be considered with estimated speed up to 400 Km/h.

*G. Security capabilities*

Security is a wide and complex field, crucial to support communication between NATO coalition partners as well as national solutions. The following issues shall be considered:

*1) INFOSEC:* the MBWCS will support up to NATO security classification level 3 (NATO SECRET or national equivalent) for big units deployments and up to level 2 (NATO CONFIDENTIAL or national equivalent) for small units. NATO coalition partners as well as national security systems with different security levels will get connected to the networks. Additionally, MBWCS will be able to switch between software or hardware-based ciphering systems.

*2) COMSEC:* the MBWCS shall adapt or use several security mechanisms based on national and coalition specific cryptographic solutions, hence supporting key management features including: Generation, Activation, Deactivation, Reactivation and Destruction of Keys and the Authentication, Authorization, and Accounting (AAA) concept. Even when critical information is secured (ciphered), the unauthorized user can act as an eavesdropper and start simple communication behavior analysis. Depending on the level of signal knowledge, the unauthorized user may act as a communication participant while attacking. To prevent the influence of such attacks, several protection mechanisms and Electronic Protection Measures (EPM) features have been identified within TRANSEC capabilities: Low Probability of Interception (LPI), Low Probability of Detection (LPD) and Anti-Jamming (AJ).

*3) NETSEC:* the MBWCS shall support protection mechanisms including incorrect traffic generation such as denial-of-service attacks (e.g. cache poisoning, message bombing), incorrect traffic relaying (e.g. blackhole, replay, wormhole and rushing attacks as well as message tampering), and error correction capabilities.

*H. Robustness capabilities*

The MBWCS will provide robustness to signal interference and/or loss of network operation. When deployed in locations with other tactical networks, i.e. vehicular deployment, it will provide adequate measures to avoid interference from adjacent users in the same frequency band. For mesh or PMP modes, the network will provide redundancy and be robust to a single point of failure. This may be of the form of a link failure or the failure of a radio, without unduly affecting the overall network performance. Systems will be robust to jamming signals in the form of noise, barrage, and sweep/chirp jamming, supporting techniques to actively track jamming signals and applying automatic jamming avoidance measures. The MBWCS should include cognitive radio and dynamic spectrum management techniques to automatically overcome bad conditions in the communications environment.

The operational requirements for robustness also include the physical attributes of the radio. Generally, this is addressed by the target platform requirements which in turn is dependent on the deployment scenario. Equipment will be physically robust to environmental damage, i.e. shock- and water-proof. The MBWCS will provide the mechanisms to allow fast switching between the technology chosen and back-up/legacy communications in the event of failure. The MBWCS will support an uninterrupted power supply to ensure that a back-up power supply can support a minimum of 15 minutes for small units and around 1-2 hours for big units; maintaining the continuous usage of the radio platform for a minimum of 3 months without interruption for big units; and in the order of magnitude of days for small units. When deployed in a handheld or manpack radio configuration, the MBWCS will have power requirements compatible with existing battery capabilities.

*I. Target frequency bands*

NATO Band IV, from 4.4 to 5 GHz, allows high throughputs enabling the usage of advanced services with smaller coverage than in HF, VHF or UHF bands. Operational concepts for NATO III+ and IV frequency bands will cover a wideband PtP and PMP radio-link at the higher level of the military echelons with no or limited mobility, hence addressing scenarios Type A and B. Typical channel bandwidth is among 10-20 MHz providing a high data rate backbone. NATO Band I, from 225 MHz to 400 MHz, and its potential migration to 1-2 GHz frequency band (part of the NATO III frequency band) is used between Battalion and Brigade level. This is still the target band for the systems that are currently being developed. This band has restrictions such as the reduction of the channelization bandwidth, nevertheless it offers the possibility of a significant increase in the range of communications. Operational concepts for NATO Band I frequency band are mainly addressing scenarios with full mobility and MANET capabilities, at Company level or below (Type D and E), with a typical channel bandwidth of 1.25 MHz and able to provide data services up to 1 Mbps together with voice services.

*J. Coverage capabilities*

In order to increase coverage and allow for higher performances, Brigade CC and its Battalion CCs as well as Companies will provide relay functionality in NLOS conditions either in suburban or in rural areas (including coastal scenarios). Mesh will be considered at least inside Company deployments. Brigade CC and its Battalion CCs will communicate with each other considering that maximum distance of one hop among them is maximum 60 km for LOS conditions (maybe with degraded performances), Battalion CC and its Companies with a maximum of 20 km and Battalion CCs with a maximum distance of one hop among them of 60 km.

TABLE II: table:Compliance Matrix of WiMAX, LTE and WLAN

| C | Requirements | WiMAX | LTE | WLAN |
|---|---|---|---|---|
| Deployment | **PHY**: Power efficient modulations. | PC | PC | NC |
| | **PHY**: Efficient coding schemes. | C | C | C |
| | **CL**: Power management with different operation modes and fast-switching technologies. | C | C | C |
| Management | **MGT**: Specific APIs based on the POSIX standard to allow the waveform to be fully reconfigured. This includes, but not limited to, the ability to change the transmission frequency, modulation and coding and network QoS. | NC | NC | NC |
| | **MGT**: An interactive system architecture, i.e. modular-view-controller architecture patterns, to reconfigure the waveform via a specific Application Programming Interface (API). | NC | NC | NC |
| | **MGT**: A collection of pre-defined parameters in an user profile to allow easy configuration and deployment based on operational scenarios. | NC | NC | NC |
| | **PHY**: Spectrum sensing or the utilization of a sensor network at physical layer as additional features to provide feedback for the system planners. | PC | PC | PC |
| Services and Applications | **CL**: The MAC layer shall support burst data traffic with high peak rate demand, simultaneously supporting streaming video and latency-sensitive voice traffic as well as other data/Web services like e-mail, chat, file/tactical data transfer over the same channel. | C | C | C |
| | **NET**: Developed for the delivery of IP-based broadband services. | C | C | C |
| | **MAC**: The transmission time interval used by MBWCS as well as MAC Layer/Scheduler shall be able to provide real-time requirements. | C | C | PC |
| | **CL**: VOIP connections. | C | PC | PC |
| | **CL**: MBWCS shall provide data latency for voice data transfer less than 300 ms, for video data transfer, at least 1 Mbps data rate and data latency less than 200 ms for the low criticality data transfer at least 9.6 Kbps data rate and data latency less than 1s for the critical data transfer at least 384 Kbps data rate and less than 200 ms data latency. | C | C | C |
| | **MAC**: A specific scheduling algorithm in order to provide the necessary QoS for time-sensitive traffic such as voice and video according to the previous technical requirements. | PC | PC | PC |
| | **NET**: Networking QoS features include: bandwidth, delay, error, availability, Security. | PC | PC | PC |
| | **CL**: Congestion management, traffic shaping and packet classification features. | C | C | C |
| | **NET**: Routing information shall take priority over any other traffic. | PC | NC | C |
| Network | **NET**: MBWCS shall support IP protocols (IPv4 / IPv6) to enable IP based NNEC concept with broadcast, multicast and unicast capabilities. | C | C | C |
| | **NET**: Connection oriented (e.g. TCP) and connectionless (e.g. UDP) services as well as applications like Session Initiation Protocol (SIP) or the IP Multimedia Subsystem (IMS) architecture. | C | C | C |
| | **NET**: Efficient IP services including several compression techniques: Packet Header Suppression (PHS), Robust Header Compression (ROHC) or Enhanced Compressed Real Time Protocol (ECRTP). | C | C | NC |
| | **MAC**: Automatic Repeat Request (ARQ) techniques (fast retransmissions). | C | C | PC |
| | **MAC**: Relay capabilities (extended range, backbone connections) to avoid communication gaps. | C | C | C |
| | **CL**: Cross-layering techniques in order to support several basic capabilities like or QoS management, shall be considered. | C | C | PC |
| | **NET**: Mobility management in the network layer (e.g. scenario Type 2, 3 and 5). This can be done by supporting mobile IP protocols like mobile IPv6, hierarchical mobile IPv6, fast mobile IPv6 or Proxy Mobile IPv6 (at network side). | C | C | PC |
| Topology | **CL**: MBWCS with MANET topology shall support dynamic network environments between vehicle convoys or groups of dismounted personnel where nodes may regularly join or leave the network and the connectivity between nodes may change frequently. | PC | NC | NC |
| | **CL**: Network protocols with ad-hoc, self-healing, self-forming and path optimizing capabilities. | PC | PC | C |
| | **NET**: Network Layer MANET routing protocol shall consider the following features in order to maximize network efficiency; distributed operating; loop-freedom (open, closed); proactive operation in case of enough bandwidth and energy supply permission, i.e. QOLSR, Fast-OLSR, TBRPF, OSPF, OLSRv2 · · · , hybrid Operation and security. | NC | NC | PC |
| | **MAC**: Mechanisms for bandwidth request and assignment. | C | C | PC |
| | **CL**: Power control and Adaptive Modulation Control (AMC) mechanisms. | C | C | C |
| | **CL**: The MAC layer shall support network entry, ranging, key management, multicast...according to the network topology. | C | PC | C |
| Mobility | **PHY**: Physical layer of MBWCS shall have an appropriate frame structure and parameters (such as reference signals, cyclic prefix, sub-carrier spacing ($L$:f), time delay imposed and so on) in order to mitigate the errors to be formed due to the Doppler Effect. | C | C | NC |
| | **MAC**: MBWCS MAC layer shall be able to establish different links at the same time for handover. | NC | NC | NC |
| | **PHY**: MBWCS PHY Layer shall be able to provide metrics, such as SINR and RSSI, to measure the link quality. | C | C | PC |

| C | Requirements | WiMAX | LTE | WLAN |
|---|---|---|---|---|
| Mobility | **MAC**: MBWCS MAC Layer shall use the provided metrics to take handover decisions. | C | C | C |
| | **CL**: MBWCS ecosystem shall provide a backbone infrastructure for mobility management signaling exchange in order to perform handover mechanisms. | C | C | U |
| Security | **SEC, TRANSEC**: Frequency hopping and spread-spectrum techniques (LPD). | NC | PC | NC |
| | **PHY, TRANSEC**: MIMO and/or smart antennas due to Direction Of Arrival (DOA) (LPD). | C | C | NC |
| | **TRANSEC**: secure PN-sequence generators to prevent easy sequence estimation (LPI). | PC | C | NC |
| | **TRANSEC**: scrambling of transmission data and control information (LPI). | PC | C | C |
| | **CRYPTOSEC**: ciphering, authentication and key management algorithms adaptable to national or coalition needs, support for NATO Suite B. | PC | U | PC |
| | **CRYPTOSEC**: mutual authentication even for non-equal treated stations, i.e. BS and SS. | C | C | PC |
| | **CRYPTOSEC**: internal and external security devices for ciphering (IPSEC: IP ciphering), digital signatures and the possibility to volatile store critical material (keys, policies, algorithms). | PC | C | NC |
| | **MGT**: MBWCS shall support Over-The-Air (OTA) operations, e.g. transmission of security material using Over-The-Air Rekeying (OTAR). | C | NC | PC |
| | **INFOSEC**: NATO Level 3 security including IP security protocols (IPSec/HAIPE) as well as IP tunneling protocols (NAT, IPv4/IPv6-Transition). | C | C | C |
| Robustness | **MAC**: Adaptive modulation and coding and/or HARQ or ARQ strategies to offer robustness to interference. | C | C | PC |
| | **PHY**: Depending on the deployment, the system shall be compliant with spectral emission masks to avoid co-site interference. | C | C | C |
| | **PHY**: The MBCWS shall employ interference cancellation techniques to mitigate the effects of jamming signals. | PC | PC | NC |
| | **CL**: Algorithms and signal processing techniques to actively track jamming signals and instantiate algorithms in both the physical and network layer, to allow the radio to signal and change certain transmission profiles such as transmission frequency. | NC | C | PC |
| | **CL**: MAC or Network layer signaling algorithms to provide sufficient channel quality indicators, thus allowing the fast adaptation of the network to interference signals. | C | C | C |
| | **NET**: In the case of a loss in an external synchronization signal i.e. GPS/GNSS, the system shall be designed to self-configure and maintain network connectivity. | C | C | C |
| | **PHY**: MIMO and/or beam-forming techniques shall be required to improve the link performance. | C | C | C |
| | **CL**: Dependent on the deployment scenario, power control algorithms and sleep and idle modes shall be provided to conserve power consumption. | C | C | C |
| | **CL**: Network self-healing and recovery, the loss of a single radio or link can not affect network performance. | C | PC | C |
| | **PHY**: Channel coding in the form of forward error correction codes shall be designed in order to increase the robustness offered by these techniques. | C | C | C |
| Bands | **PHY**: A specific profile designed for NATO I target frequency band, based on limited bandwidths (i.e., 1.25 MHz) and single carrier modulations. | NC | NC | NC |
| | **PHY**: A specific profile designed for NATO IV target frequency band, based on large bandwidths (i.e., 20 MHz or higher) and multicarrier modulations. | PC | NC | NC |
| Coverage | **PHY**: MBWCS shall be able to establish links in both LOS and Non Line-Of-Sight (NLOS). | C | C | PC |
| | **CL**: MBWCS shall support Layer 2 or Layer 3 Relay technology in order to extend coverage. | C | C | C |
| | **CL**: The Physical Layer as well as MAC layer of MBWCS shall be in accordance with Relay Technology used. | C | C | C |
| | **CL**: Mesh Networking for Company Level communication being capable to route or switch through the traffic of other nodes in order to extend coverage. | C | PC | C |
| | **PHY**: MBWCS shall be able to assign a lower frequency channel with low data rate option (changing to a more robust modulation scheme) in order to increase coverage between two nodes without using intermediate network nodes. | NC | NC | PC |
| | **PHY**: MBWCS shall be tested according to ITU (International Telecommunication Union) Channel Models (ITU-R recommendation M.1225 and IMT-Advanced M.2135-1 (2009)) for suburban, rural and costal scenarios. | NC | NC | NC |
| Interoperability | **CL**: MBWCS shall support interoperability due to waveform concept and definition. This means a clarified definition of PHY, MAC and NET functionality and behavior and additional physical issues e.g. propagation towards routing or a definition of a common set of transmission protocols. | C | C | C |
| | **NET**: MBWCS shall support IPv4 and IPv6 protocols to enable upper layer protocols and applications. | C | C | C |
| Target | **PHY**: RF front-ends of the fixed, vehicular and man-pack platforms shall support MIMO technology. Hand-held configurations can consider as optional the support of MIMO technology. | C | C | C |

*K. Interoperability capabilities*

MBWCS will be fully compliant with NATO Reference Architecture and national-wide standards. MBWCS will be compact, reprogrammable and multi-mode, thus providing interoperability on the air by the usage of common waveforms.

*L. Target platforms*

According to the scenarios' definition, several objective platforms will be considered. Target vehicle platform will support operations on land vehicles, war ships or helicopters acting as support of the network. Nevertheless, specific platforms could operate as fixed installations like headquarters in certain scenarios (mainly Battalion, Brigade or upper levels), or hand-held or man-pack platforms at a lower tactical level (mainly companies and platoons). Deployment features and environmental conditions previously explained will be considered.

## V. An overview of the methodology

This research was conducted following a scenario-based layerized approach allocating technical requirements in the involved OSI layers. Cross-layering (CL) is used when several layers are affected simultaneously. Standards compliance and modifications' identification were assessed for each of these layers considering both waveform (WF) and platform (PTF) requirements; concluding whether the functionality can be directly derived from the standards as they are or if, at a high level, modifications are needed. This structure optimizes the comparison between different standards, helping in the definition of the final MBWCS proposal. A cost-benefit analysis of the implementation of the modifications of each one of the standards was performed. The aim is to provide some qualitative metrics about the effort needed for conducting these modifications against the benefits/impact achieved in terms of compliance. In summary, a compliance matrix for technical requirements shows the analysis result with the criteria fully or partially compliant or not. This matrix is essential as guidance for the specification of the ideal MBWCS. The cost-benefit analysis of the implementation of the modifications of WiMAX, LTE and WLAN, and the specifics of scenarios A, B, C, D and E are also considered to structure analysis' outcomes. In order to simplify, this survey does not go into detail of each one of the scenarios' issues.

## VI. Applicability analysis

The aim of this section is to shortly describe the applicability analysis of the targeted standards confronting the identified technical requirements.

*A. PTF Requirements*

Following, some of the PTF-only requirements are cited: reduced weight and dimension equipment, with the highest level of integration, ease of installation and plug and play (Portable platforms: man-pack with size 257 cu. in. (438 cu. in. with battery), maximum 3" H × 10" W × 9" D (without battery bucket), 3" H × 10" W × 14" D (with battery bucket), weight 9 lbs. (14 lbs. with battery) and hand-held with size 28 cu. in. and weight 1.7 lbs with battery and antenna; Vehicular platforms: 5.472" / 7.67" H × 11.4" / 15.74" W × 12.59" / 13.38" D). Antennas shall be carefully chosen considering deployment type scenario: fixed/vehicular/man-pack/hand-held, external/internal location and height consistent with coverage range (according to free Fresnel zone), polarization, beamwidth, gain ... Omnidirectional antennas shall be chosen when high mobility is required (scenarios Type C, D and E) along with the incorporation of features like auto-acquisition, optimum orientation, tracking ...

A default codec for narrowband voice (such as G.711, G.726, G.729AB and G.723.1), a default codec for wideband voice (such as G.722, G.722.2) and a default codec for fax (such as G.711) for VOIP service. The MBWCS shall provide 28.8-87.2 Kbps data bandwidth depending on chosen CODEC, for narrowband voice (such as G.711, G.726, G.729AB and G.723.1), wideband voice or VOIP service.

An SNMP/HTTP based network management is needed to support remote network management. The network shall be configured, in all the elements of the architecture: BS, CPE and backbone, to provide redundancy in such a way that no single loss of a node will result in the degradation of services or loss in communications. ARP protocol for connections with external networks (Ethernet) and special mechanisms, e.g. gratuitous ARP, are needed together with systems for avoiding intrusion and/or tampering, e.g. firewalls, anti-virus software or malware scanners.

Procedures, design values and equipment shall be compliant with the considerations from military standards: MIL-STD 810G, MIL-STD 461F, MIL-STD-1275 ...

The MBWCS shall supply a common interface (connectors to radios and software) to support possible external crypto modules and a FILL interface for security material handling. Tunable hardware filters at the receiver front-end with variable bandwidths will be needed to accommodate the various modes of operation to avoid co-site interference. The MBWCS shall provide a GPS antenna interface and embedded GPS receiver to support synchronization capabilities.

The platform shall provide specific physical interfaces; for example, for control purposes, control interfaces can be mapped on a RS-232 or Ethernet interface. For payload transmission and reception, interfaces can be mapped on an Ethernet interface. For voice communications, interfaces can be mapped on a PTT interface, Ethernet or any other specific interface.

*B. WF Requirements*

The set of 4G standards, as can be seen in Table II, covers the main necessities identified in terms of advanced services support with enough QoS and mobility support, mainly having gaps in their adaptation to specific military frequency bands, security, and robustness.

Specifically, WiMAX, LTE and WLAN are compliant with WF or WF/PTF requirements such as: efficient coding schemes, power management with different operation modes and fast-switching technologies, congestion management, traffic shaping and packet classification features, power control, AMC mechanisms and relay capabilities, MIMO and/or beam-forming techniques ...

Their MAC layer support burst data traffic with high peak rate demand, simultaneously supporting streaming video and latency-sensitive voice traffic as well as other data/Web services. The 4G MBWCS provides data latency for voice data transfer less than 300 ms; for video data transfer, at least 1 Mbps data rate and data latency less than 200 ms; for the low criticality data transfer at least 9.6 Kbps data rate and data latency less than 1s; for the critical data transfer at least 384 Kbps data rate and data latency less than 200 ms.

The standards support IPv4 / IPv6 to enable IP based NNEC concept with broadcast, multicast and unicast capabilities, connection oriented (TCP) and connectionless (UDP) services as well as applications like SIP or the IMS architecture. NATO Level 3 security including IP security protocols (IPSec/HAIPE) as well as IP tunneling protocols (NAT, IPv4/IPv6-Transition) are supported.

In other requirements WiMAX, LTE and WLAN just partially comply, for example in spectrum sensing or the utilization of a sensor network at physical layer as additional features to provide feedback for the system planners. The standards also do not present the ideal scheduling algorithm in order to provide the necessary QoS for time-sensitive traffic such as voice.

WLAN is the only standard that is partially or non-compliant with the transmission time interval as well as MAC Layer/Scheduler real-time requirements, efficient IP services including several compression techniques: PHS, ROHC or ECRTP, adaptive modulation and coding and/or HARQ or ARQ strategies to offer robustness to interference, cross-layering techniques in order to support several basic capabilities like or QoS management, mobility management in the network layer by supporting mobile IP protocols like mobile IPv6, hierarchical mobile IPv6, fast mobile IPv6 or Proxy Mobile IPv6 (at network side), among others.

Nevertheless, none of the standards completely fulfill the following requirements: an interactive system architecture, for example modular-view-controller architecture patterns, which allow for reconfigurability of the waveform via a specific API. These standards do not consider a collection of pre-defined parameters in a user profile to allow easy configuration and deployment based on operational scenarios, and a MAC layer able to establish different links at the same time for handover.

The interoperability due to waveform concept and definition means a clarified definition of PHY, MAC and NET functionality and behavior, and additional physical issues considerations e.g. propagation towards routing or a definition of a common set of transmission protocols.

The PHY layer design is clearly driven by TRANSEC features, and is significantly different of an OFDM-based system with high bandwidth efficient modulation, this implies the implementation of power efficient modulations, or frequency hopping and spread-spectrum techniques. Physical layer of MBWCS shall have an appropriate frame structure and parameters (such as reference signals, cyclic prefix, sub-carrier spacing ($L$: f), time delay imposed and so on) in order to mitigate the errors to be formed due to the Doppler Effect, and efficient techniques and/or algorithms in order to reduce PAPR in downlink path.

Only LTE offers secure PN-sequence generators to prevent easy sequence estimation, scrambling of information and support for CRYPTOSEC capabilities: internal and external security devices for ciphering (IPSEC: IP ciphering), digital signatures and the possibility to volatile store critical material (keys, policies, algorithms). Although WiMAX, in exclusive, gives support for Over-The-Air (OTA) operations, e.g. transmission of security material using OTA Rekeying (OTAR) or is capable to route or switch through the traffic of other nodes in order to extend coverage in mesh networking for Company Level communications.

None of them have specific profiles designed for NATO I and IV and they need improved protocol stacks for supporting MANET topologies considering hybrid operation and security features.

The definition of MBWCS merges the most promising and compliant components or blocks according to these outcomes to reach its full potential.

## VII. Conclusions

This section summarizes the results of the study and details the most immediate milestones. It was proved that the development of an innovative MBWCS would be clearly optimized if 4G standards are taken as basis. Once the feasibility has been confirmed, and after a technical and cost/benefit analysis of the implementation of a 4G scenario-based MBWCS, the way-ahead is the setting up of a specific Research Task Group (RTG) for NATO IST-ET-068. This RTG should evolve the high-level assessment into the quantitative domain, thus performing a detailed design of the envisaged MBWCS, conducting exhaustive simulations and prototyping activities with the WiMAX, LTE and WLAN promising features and modules concerning the specified requirements' compliance. Identified enabling technologies, such as cognitive radio shall be examined. Conclusions state that standards only imply a partial compliance of some of the requirements identified and none of them are able to comply with the full specification, although this work gives an overall view of the most efficient and timely way to design a MBWCS for the near future warfare.

## Acknowledgment

This work has been funded by the GTEC (Group of Electronic Technology and Communications) of the University of A Coruña and Indra Sistemas S.A. The authors acknowledge to Colin Brown, Mehmet Hayri Küçüktabak and Matthias Tschauner their collaboration in the NATO IST-ET-068.